\documentclass[aps,twocolumn,showpacs,preprintnumbers,amsmath,amssymb,superscriptaddress,balancelastpage,prb]{revtex4-2}

\include{options}
\usepackage {amsmath,amssymb,epsfig,color,ulem}
\usepackage{natbib}
\usepackage {graphicx}

\begin{document}

\title{Evolution of the pseudogap band structure in a system of electron-correlated lattice polarons}
\author{E.~I.~Shneyder}
\email{eshneyder@gmail.com}
\affiliation{Kirensky Institute of Physics, Federal Research Center KSC SB RAS, 660036 Krasnoyarsk, Russia}
\author{M.~V.~Zotova}
\affiliation{Kirensky Institute of Physics, Federal Research Center KSC SB RAS, 660036 Krasnoyarsk, Russia}
\affiliation{Siberian Federal University, 660041 Krasnoyarsk, Russia}
\author{A.~V.~Dudarev}
\affiliation{Kirensky Institute of Physics, Federal Research Center KSC SB RAS, 660036 Krasnoyarsk, Russia}
\affiliation{Reshetnev Siberian State University of Science and Technology, 660037, Krasnoyarsk, Russia}
\author{S.~V.~Nikolaev}
\affiliation{Kirensky Institute of Physics, Federal Research Center KSC SB RAS, 660036 Krasnoyarsk, Russia}
\affiliation{Siberian Federal University, 660041 Krasnoyarsk, Russia}
\author{S.~G.~Ovchinnikov}
\affiliation{Kirensky Institute of Physics, Federal Research Center KSC SB RAS, 660036 Krasnoyarsk, Russia}
\affiliation{Siberian Federal University, 660041 Krasnoyarsk, Russia}
\date{\today}

%===========================================================================
\begin {abstract}
The evolution of the role of lattice vibrations in the formation of the pseudogap state in strongly correlated electron systems has been investigated concerning changes in the electron-phonon coupling parameters and the concentration of doped charge carriers. We apply the polaronic version of the generalized tight-binding method to analyze the band structure of a realistic multiband two-dimensional model that incorporates the electron-lattice contributions of both Holstein and Peierls types. It has been demonstrated that the emergence of polaronic effects begins with the modulation of spectral function intensity. However, within a specific region of the phase diagram, a significant transformation of the electron band structure and pseudogap state occurs. It results from coherent polaron excitations that create a partially flat band near the Fermi level. This process leads to a change in the topology of the Fermi surface and the emergence of corresponding features in the density of states.
\end {abstract}

%===========================================================================
% 71.38.-k Polarons and electron-phonon interactions
% 71.27.+a Strongly correlated electron systems; heavy fermions
% 63.20.Ls Phonon interactions with other quasiparticles
% 71.10.Fd Lattice fermion models (Hubbard model, etc.)
% 74.72.-h Cuprate superconductors
\maketitle

%===========================================================================
\section{Introduction \label{intro}}
%===========================================================================
The term "pseudogap" was introduced by Nevill Mott to describe the reduction in the density of electronic states at the Fermi level, which arises due to Coulomb repulsion between electrons at the lattice sites, the formation of a forbidden energy gap in disordered systems, or a combination of these factors~\cite{RevModPhys.40.677}. Pseudogap anomalies serve as distinctive features of systems with strong electronic correlations that are near instabilities, leading to new forms of order, phase transitions, or quantum phase transitions. Since the discovery of high-temperature superconductors (HTSCs) based on complex copper oxides, significant attention has been devoted by researchers to the pseudogap effects in the electronic structure of HTSC materials. These effects, observed in doped compounds below a critical temperature $T^*$, manifest as a decrease in the spectral weight of quasiparticle excitations in the vicinity of the chemical potential. Current issues relevant to understanding the nature of HTSCs regarding what the principal mechanism is for the formation of the pseudogap in cuprates, whether the pseudogap phase precedes or competes with the superconducting phase, remain under discussion~\cite{Sadovskii_2001,Sato2017,PhysRevX.13.031010,Vedeneev_2021,PhysRevB.110.024521,doi:10.1126/sciadv.adg9211}.

Like many other compounds exhibiting pseudogap anomalies, cuprates demonstrate a complex phase diagram characterized by the competition among Coulomb, exchange, and electron-boson interaction effects. Among these factors, the electron-phonon interaction (EPI) plays a considerable role. The significant involvement of lattice vibrations in the formation of the pseudogap state is confirmed by substantial changes in temperature $T^*$ observed during isotopic substitution of oxygen~\cite{Lanzara_1999,PhysRevLett.84.1990,PhysRevB.66.184506,PhysRevB.74.184520,PhysRevB.95.014514} and copper~\cite{EurPhysJB.19.5} atoms, as well as by the temperature-dependent evolution of the frequency and linewidth of individual phonon modes, which correlate with the emergence of the pseudogap~\cite{PhysRevB.107.224508}. The study of electron-lattice interaction effects in the processes of modulation and formation of pseudogap states within systems characterized by strong electron correlations is of great interest for a wide range of compounds, especially those containing metals with partially filled d- or f-orbitals~\cite{Rini2007,Wang2023}. Recently, this topic has garnered considerable interest from researchers seeking methods to modify the properties of correlated systems~\cite{Bukharaev_2018,Ramesh2019}, especially through light-induced phonon excitations~\cite{Cavalleri2011,Valmispild2024}.

From a theoretical viewpoint, the most thoroughly investigated model is Hubbard-Holstein at half-filling. In the non-adiabatic regime, characterized by phonon mode frequencies ${{\omega }_{ph}}$ significantly exceeding the Fermi energy $\omega_F$, this scenario is the least resource-intensive in terms of computational effort. Conversely, the ratio $ {{\omega }_{ph}} \ll \omega_F $, which is typical for many compounds, leads to challenges associated with the rapid expansion of the Hilbert space of states as the strength of electron-phonon coupling increases. On the one hand, the present development of numerical and analytical approaches for studying systems with strong electronic correlations allows for a fairly reliable description of the characteristics of their ground state, as well as their spectroscopic, transport, and nonequilibrium properties~\cite{Rubtsov2023,Ridley2022,PhysRevB.71.125119,PhysRevLett.93.076401}. On the other hand, the inclusion of electron-boson degrees of freedom critically raises the demands on computational resources, complicating the application of modern methods across a wide range of parameters~\cite{PhysRevB.72.035122,Fehske2007,PhysRevB.77.125101,PhysRevB.82.085116,PhysRevLett.113.166402,PhysRevB.92.155143,PhysRevResearch.2.043258,PhysRevB.104.035106,PhysRevB.103.115123}.

The addition of strong electron-phonon couplings to a system with strong electron correlations results in a modulation of the competition between the kinetic energy of the carriers and their Coulomb interaction energy, all set against the backdrop of lattice polaron effects. The intricate interplay of interactions gives rise to a region in the phase diagram where a unique type of quasiparticle exists~-- electron-correlated lattice polarons. The properties and conditions for their formation differ significantly from those of polarons in uncorrelated systems and are, among other things, influenced by the competition between various types of electron-phonon coupling mechanisms~\cite{PhysRevB.101.235114,PhysRevB.104.155153}. The first one is associated with changes in the local energy of charge carriers and is described by the Holstein Hamiltonian. The second corresponds to the modulation of the hopping parameters and is usually referred to as the Peierls contribution or the transitive electron-phonon interaction. Notably, the latter type of electron-lattice coupling has drawn significant attention since the seminal papers of Bari\u{s}i\'{c}, Labb\'{e}, and Friedel on transition-metal superconductivity~\cite {PhysRevLett.25.919, PhysRevB.5.932, PhysRevB.5.941} and later by Su, Schrieffer, and Heeger on soliton formation in conducting polymers~\cite{PhysRevLett.42.1698}. A proper description of the phase diagram of a polaron system requires taking both types of microscopic mechanisms of electron-phonon coupling into account. This conclusion is supported by previous studies addressing a range of issues, including the non-analytic behavior of polaron characteristics~\cite{RevModPhys.63.63,PhysRevB.78.214301,PhysRevLett.105.266605,Sboychakov2010,PhysRevB.92.155143}, crossovers between polaron and bipolaron transformations~\cite{PhysRevB.101.235114}, nontrivial topological effects~\cite{PhysRevLett.42.1698,PhysRevB.100.075126}, and antiferromagnetic order induced by transitive electron-phonon interaction~\cite{PhysRevLett.127.247203,PhysRevB.105.085151,PhysRevB.106.L081115}.

Here, we examine the evolution of spectral functions in response to variations in doping and the strength of electron-phonon coupling. We employ a realistic multiband model of a two-dimensional electron-correlated system that accounts for the lattice vibrations of both the Holstein and the Peierls types. By utilizing the concept of competition between polaronic and bipolaronic transformations within a system characterized by strong electron correlations, we provide a systematic investigation of the phase diagram. For the first time, to our knowledge, we demonstrate how the role of lattice vibrations evolves from the modulation of the pseudogap properties of the Fermi surface, caused by fluctuations in short-range antiferromagnetic order, to the emergence of a new mechanism driven by coherent, strongly bound polaron excitations that modify the band structure near the Fermi level.

The paper is organized as follows: Section~\ref{sec_model} presents the model and approaches employed in this study. In Section ~\ref{sec_PhaseDiagram}, we analytically identify the specific points of the phase diagram and compare these results with numerical analysis data. Section~\ref{sec_SpectrFunc} contains the key findings and addresses the evolution of the band structure in critical regions of the phase diagram as the parameters of electron-phonon interaction and doping are changed. Furthermore, the nature of flat band formation within the system of electron correlated lattice polarons is discussed, and the conclusions are presented in Section~\ref{sec_Discussion}.

%===========================================================================
\section{Model and calculation method \label{sec_model}}
%===========================================================================
The total Hamiltonian has the form of the extended Emery model~\cite{Shneyder18}. The analysis of electron-lattice effects is performed in the adiabatic limit, taking into account only one optical dispersionless mode, which is assumed to interact most strongly with the charge carriers~\cite{Anzai2017}.
\begin{eqnarray}
\label{H_tot}
& {H_{el}} = \sum\limits_{{\bf{g}},l,\sigma } {\left[ { \sum\limits_{\alpha }{\left( {\varepsilon _\alpha }n_{{{\bf{g}}_\alpha },\sigma }^\alpha  + \frac{1}{2}{U_\alpha}n_{{{\bf{g}}_\alpha },\sigma }^\alpha n_{{{\bf{g}}_\alpha },\bar \sigma }^\alpha \right)} + } \right.} \nonumber \\
& + \sum\limits_{{\bf{g}}',l' } {{P_{pp}}{t_{pp}}\left( {p_{{{\bf{g}}_l},\sigma }^\dag {p_{{{\bf{g}}'_{l'}},\sigma }} + {\rm{H}.c.}} \right)}  + \nonumber \\
&\left. { + {P_{pd}}{t_{pd}}\left( {d_{{\bf{g}},\sigma }^\dag {p_{{{\bf{g}}_l},\sigma }} + {\rm{H.c.}}} \right) + \sum\limits_{\sigma '} {{V_{pd}}n_{{{\bf{g}}_l},\sigma }^pn_{{\bf{g}},\sigma '}^d} } \right] + \nonumber \\
& + \sum\limits_{{\bf{g}},\sigma } {{M_d}\left( {f_{\bf{g}}^\dag  + {f_{\bf{g}}}} \right)d_{{\bf{g}},\sigma }^\dag {d_{{\bf{g}},\sigma }}}  + \nonumber \\
& + \sum\limits_{{\bf{g}},l,\sigma } {{M_{pd}}{P_{pd}}\left( {f_{\bf{g}}^\dag  + {f_{\bf{g}}}} \right)\left( {d_{{\bf{g}},\sigma }^\dag {p_{{\bf{g}}_{l},\sigma }} + {\rm{H.c.}}} \right)} \nonumber \\
& +\sum\limits_{\mathbf{g}}{\hbar {{\omega }_{ph}}\left( f_{\mathbf{g}}^{\dagger }{{f}_{\mathbf{g}}}+\frac{1}{2} \right)}
\end{eqnarray}
Here, the vector $\bf{g}$ enumerates the sites of a square lattice centered on the atoms with partially filled $d$-orbitals, $ \alpha $ is the orbital index, and $ \alpha = p,d $, with $p$ corresponding to the ligand atoms, ${\bf{g}}_\alpha = {\bf{g}}_l$ for $ \alpha= p $, where vector ${\bf{g}}_{l}$ runs through the positions of the ligand atoms in the unit cell. The values $ \varepsilon _\alpha $ determine the local atomic energies with respect to the chemical potential $\mu$. The operators $d_{\bf{g},\sigma}^\dag $ and $p_{{\bf{g}}_l,\sigma}^\dag$ create a hole with spin $\sigma$ in the $d$- or $p$-orbitals, respectively, ${n_{{\bf{g}},\sigma }^\alpha}$ are the particle number operators, $f_{\bf{g}}^\dag $ and $ f_{\bf{g}} $ are the operators of creation and annihilation of the breathing planar mode with frequency ${\omega _{ph}}$. The parameters ${U_{\alpha}}$ and ${V_{pd}}$ are the matrix elements of Coulomb repulsion, ${t_{pd}}$ and ${t_{pp}}$ are the overlap integrals of corresponding orbitals, $ - \sigma  = \bar \sigma $, $P_{pp}$ and $P_{pd}$ are the phase factors equal to either $1$  or $-1$, depending on whether the orbitals with real wave functions have the same or opposite signs in the overlap region.

Calculations were performed with the following set of parameters: ${\varepsilon _d} = 0$,  ${\varepsilon _p} = 1.5$,  ${t_{pp}} = 0.86$, $t_{pd} = 1.36$, ${U_d} = 9$, $ U_p = 4$, ${V_{pd}} = 1.5$, $\hbar \omega_{ph}  = 0.09$, and $W=2.15$ (all in eV). Therefore, the electronic part of the Hamiltonian~(\ref{H_tot}) describes a Mott-Hubbard system with charge transfer. Note that this set of parameters is typical for single-layer cuprates, such as La$_{2-x}$Sr$_x$CuO$_4$. In this paper, the case of hole doping is presented, with a concentration of doped charge carries denoted as $x$. We consider the linear in atomic displacements part of electron-phonon contribution, which stems from the modulations of the on-site energy $\varepsilon _d$ and the hopping parameter $t_{pd}$. The corresponding matrix elements are ${M_d}$ and ${M_{pd}}$. For convenience, we introduce the dimensionless parameters that characterize the strength of the electron-lattice interaction:
${\lambda _{H\left( {P} \right)}} = {{M_{d\left( {pd} \right)}^2} \mathord{\left/ {\vphantom {{M_{d\left( {pd} \right)}^2} {W\hbar \omega_{ph} }}} \right.  \kern-\nulldelimiterspace} {W\hbar \omega_{ph} }}$. Here, $W$ is the width of the valence band of electrons defined by electronic part of Hamiltonian~(\ref{H_tot}), which does not take the electron-phonon interaction into account. The indices $H$ and $P$ refer respectively to the Holstein and Peierls mechanisms of electron-phonon coupling.

The results presented below have been obtained within the polaronic version~\cite{PhysRevB.92.155143,Shneyder18} of the generalized tight-binding method (pGTB)~\cite{Ovchinnikov2012}, which is essentially a variant of the cluster perturbation theory. A distinctive feature of the approach is the introduction of an adequate set of quasiparticle excitations that are formed by simultaneously accounting for both short-range electron-electron correlations and electron-lattice coupling within an effective cluster. This makes it possible to model solutions across a fairly wide range of electron-lattice coupling parameters in a correlated system. Following the ideology of the GTB method, we carry out canonical transformations of the Hamiltonian, which allows us to separate equation~(\ref{H_tot}) into intra-cluster and inter-cluster contributions. Through exact diagonalization of the cluster Hamiltonian, we obtain its many-body eigenstates. The interactions between clusters that lead to the formation of dispersion relations for quasiparticle excitations are then analyzed using perturbation theory. To consider the inter-cluster contributions, we employ projection techniques in the equations of motion for thermodynamic two-time Green's functions~\cite{Tserkovnikov1981,Plakida2012,PhysRevB.101.235114} within a generalized Hartree-Fock approximation that takes into account the interaction between charge carriers and spin fluctuations.

\begin{figure}
\center
\includegraphics[width=0.9\linewidth]{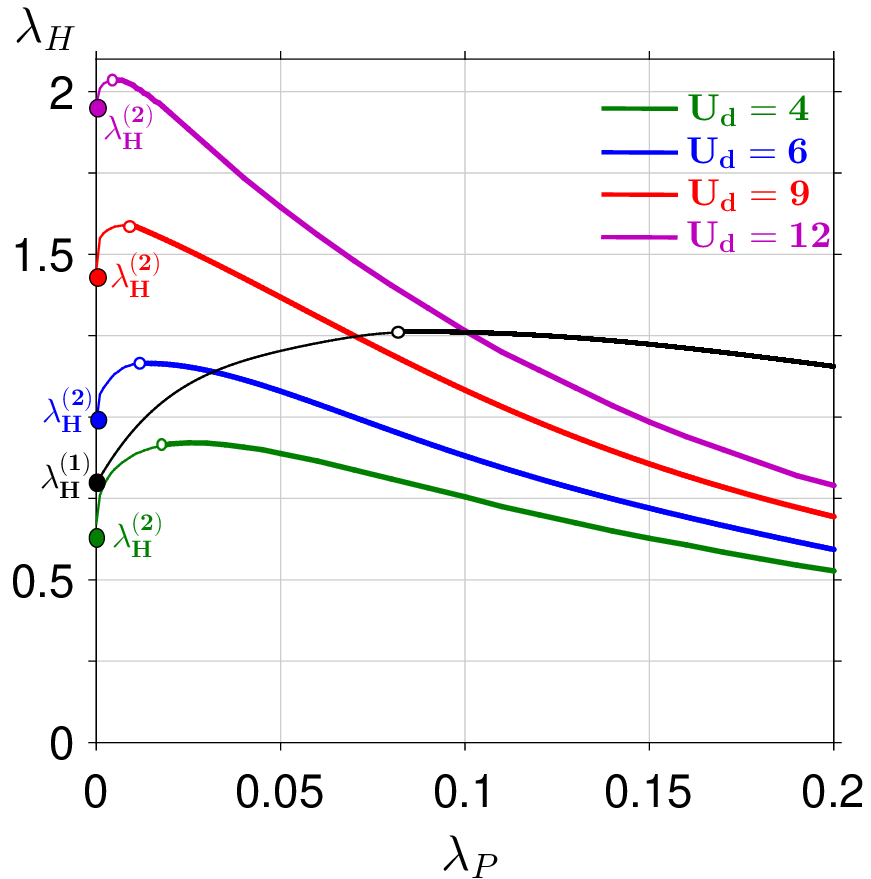}
\caption{\label{fig:f1}
Critical points and lines of the polaron (black curve) and the bipolaron (colored curves) transformations on the phase diagram of the system in the electron-lattice coupling parameters corresponding to the Holstein type $\lambda_{H}$ and the Peierls type $\lambda_{P}$. The bipolaron transformations lines depend on the values of the Coulomb repulsion $U_d$ and illustrate the collapse of the correlated polaron region in the limit of an uncorrelated or weakly correlated system, specifically when $U_d \leq 2W$. The thin and thick parts of the transformation lines, separated for convenience by unpainted circles, correspond respectively to continuous or abrupt changes in the characteristics of the bound states (more details in the text).}
\end{figure}
\begin{figure*}
\center
\includegraphics[width=0.95\linewidth]{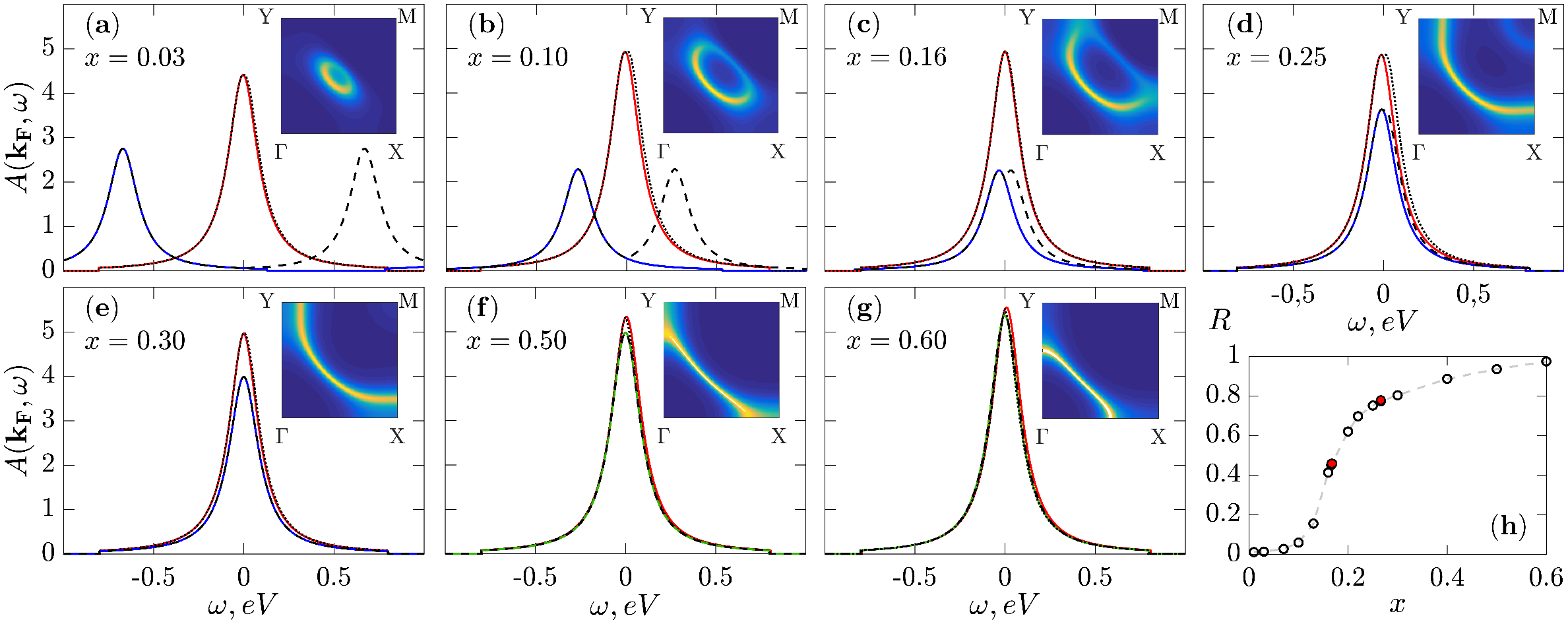}
\caption{\label{fig:f2}
Correlated system without electron-phonon interaction, $ \lambda_H = \lambda_P = 0 $: (a-g) spectral functions in the nodal direction $ \Gamma \left( {0,0} \right) \rightarrow M \left({\pi},{\pi }\right) $ (red curve) and in the antinodal directions $ X \left( {\pi,0} \right) \rightarrow M \left({\pi},{\pi }\right)$ (blue curve) and $ \Gamma \left( {0,0} \right) \rightarrow X \left({\pi},{0}\right)$ (green curve) of the Fermi surface, $\omega_F=0$; the corresponding symmetrized spectral functions are shown as dashed lines. (h) The ratio $R$ of the spectral weight intensities of the antinodal to the nodal quasiparticle peaks at the Fermi level. For images below that demonstrate the evolution of the electronic structure considering polaronic effects, only the parameters of electron-lattice coupling are specified.}
\end{figure*}
\begin{figure*}
\center
\includegraphics[width=0.95\linewidth]{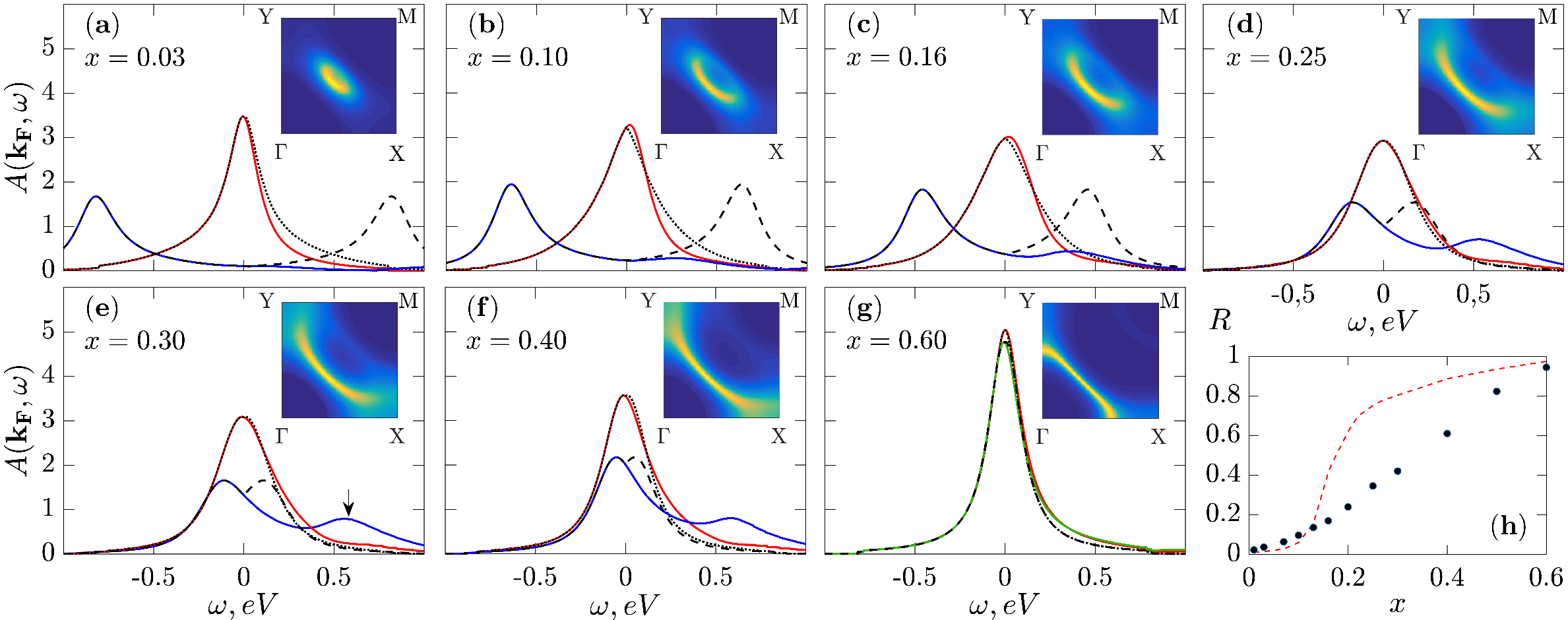}
\caption{\label{fig:f3}
$\lambda_H=0$, $\lambda_P=0.05$.
Here and below, the small black arrow indicates the emergence of a narrow polaron band at intermediate or strong electron-phonon coupling strength, (h) the ratios of the spectral weight intensities are shown with (black dots) and without (red dashed line) taking electron-phonon interaction into account.}
\end{figure*}
\begin{figure*}
\center
\includegraphics[width=0.95\linewidth]{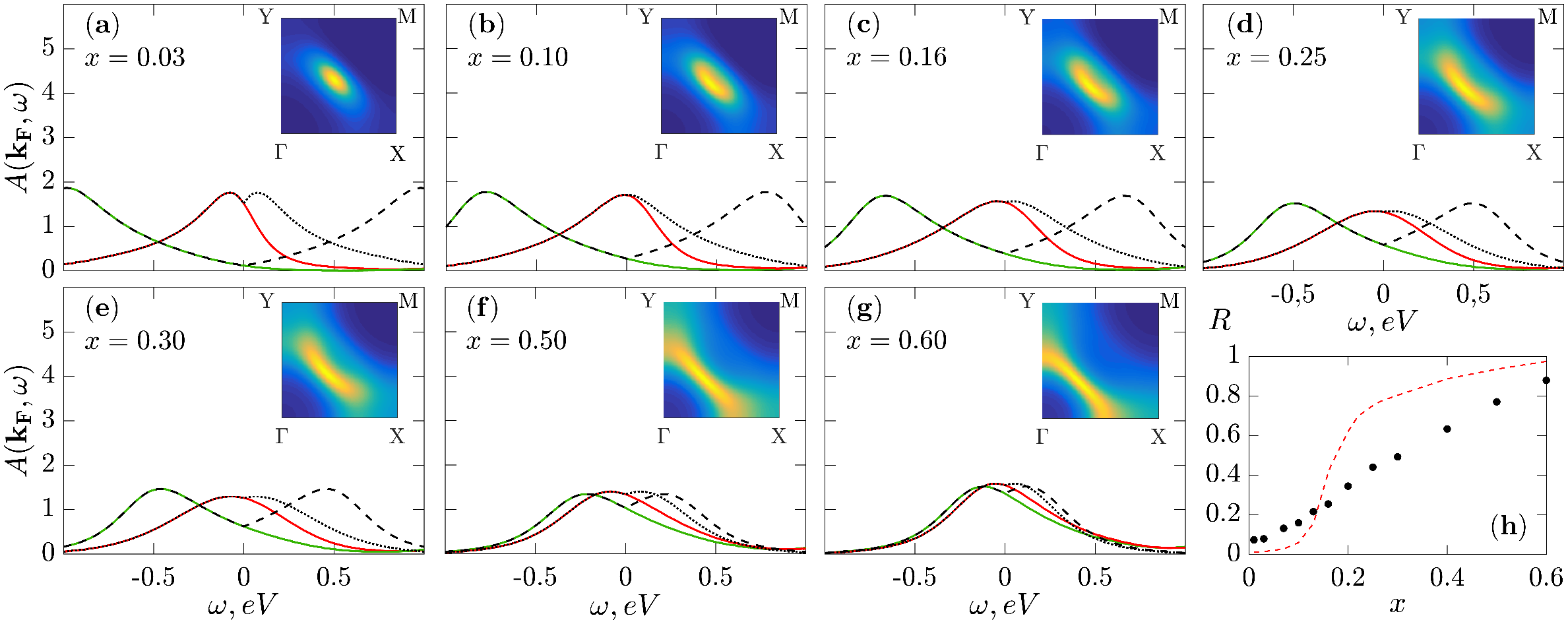}
\caption{\label{fig:f4}
$\lambda_H=0$, $\lambda_P=0.15$.}
\end{figure*}
\begin{figure*}
\center
\includegraphics[width=0.95\linewidth]{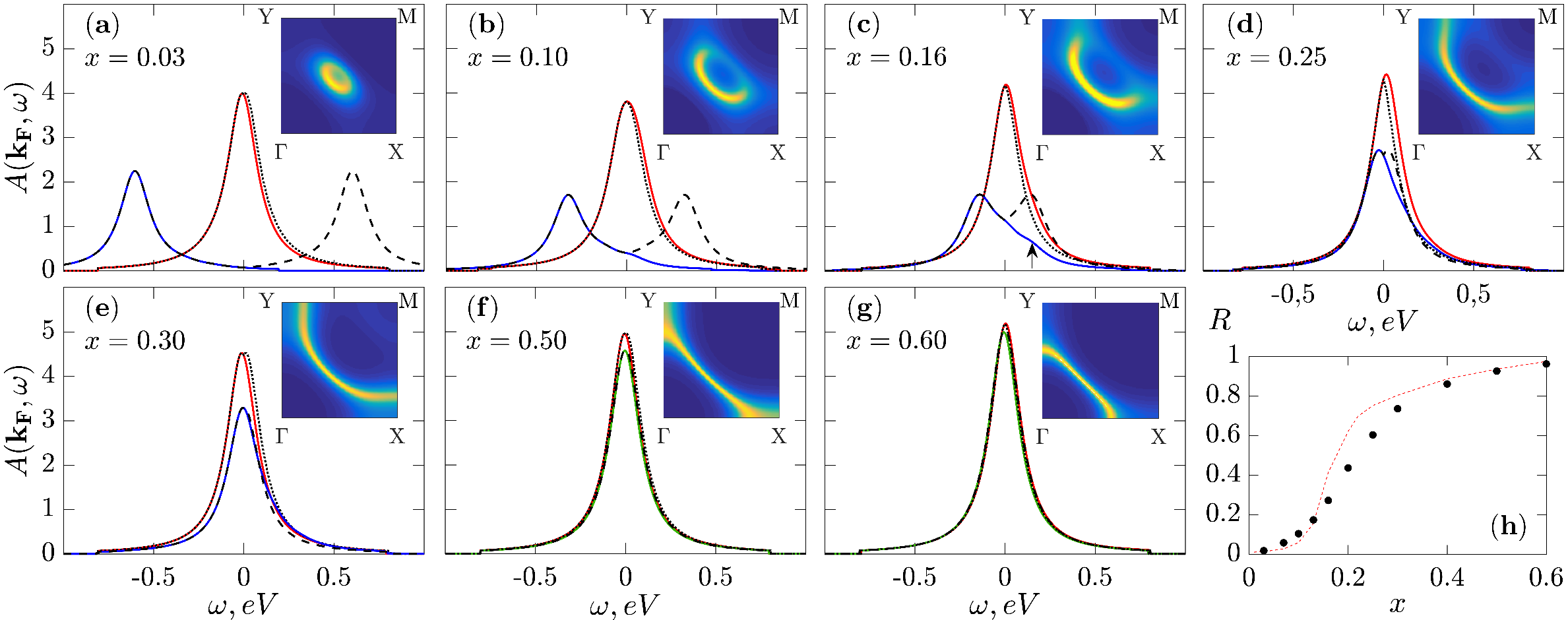}
\caption{\label{fig:f5}
$\lambda_H=0.5$, $\lambda_P=0$.}
\end{figure*}
\begin{figure*}
\center
\includegraphics[width=0.95\linewidth]{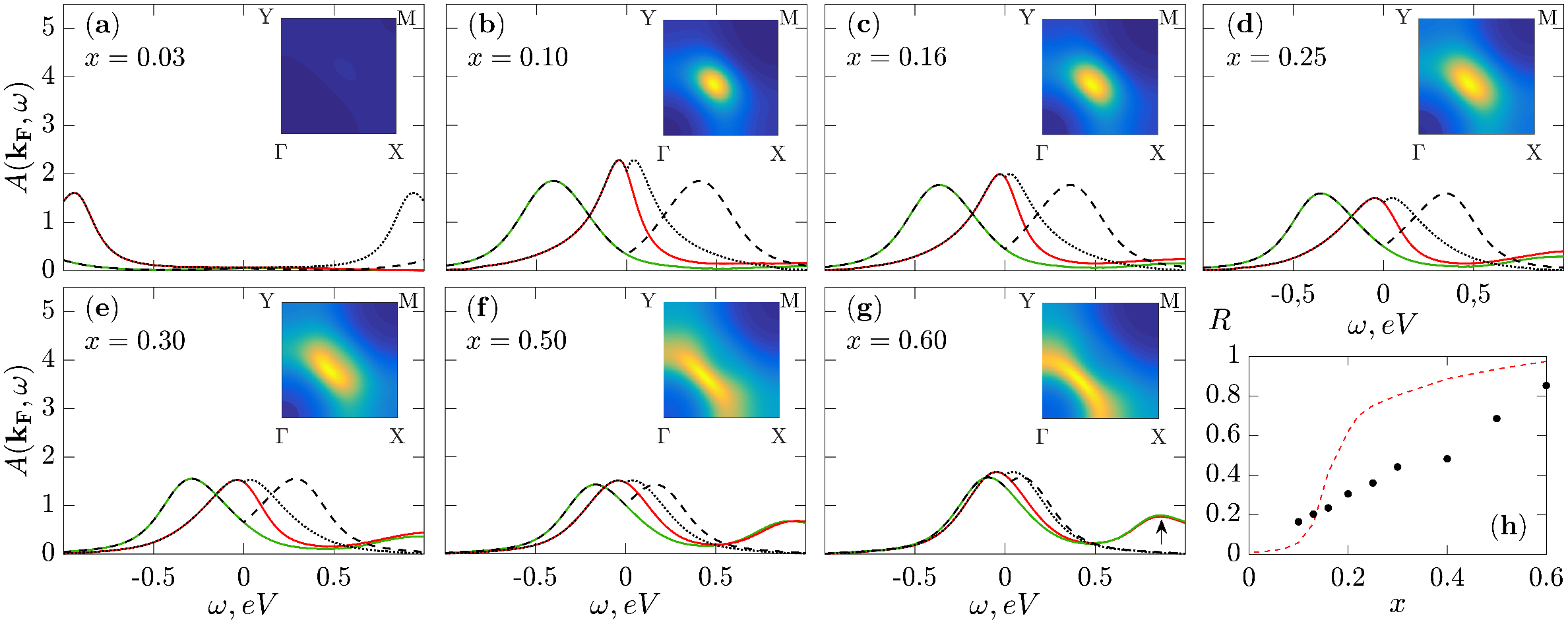}
\caption{\label{fig:f6}
$\lambda_H=1.5$, $\lambda_P=0$.}
\end{figure*}
\begin{figure*}
\center
\includegraphics[width=0.95\linewidth]{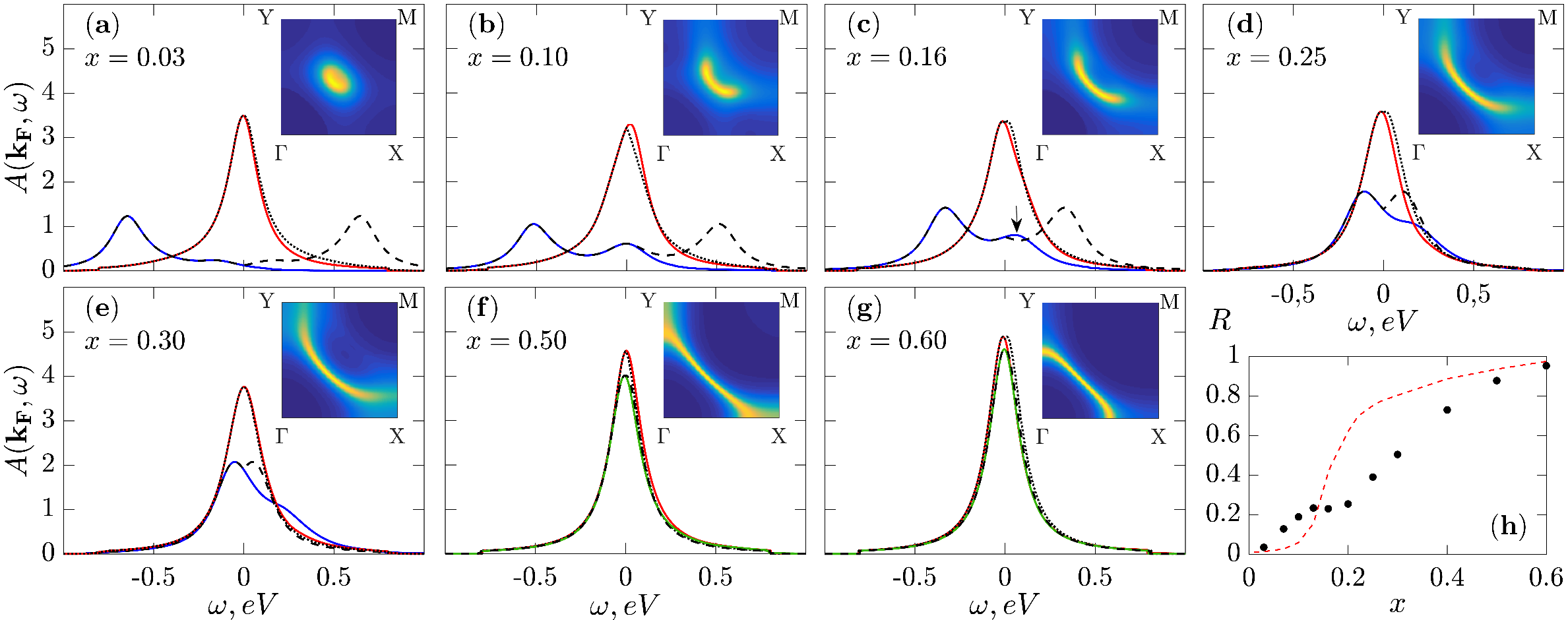}
\caption{\label{fig:f7}
$\lambda_H=1.0$, $\lambda_P=0.02$.}
\end{figure*}
\begin{figure*}
\center
\includegraphics[width=0.95\linewidth]{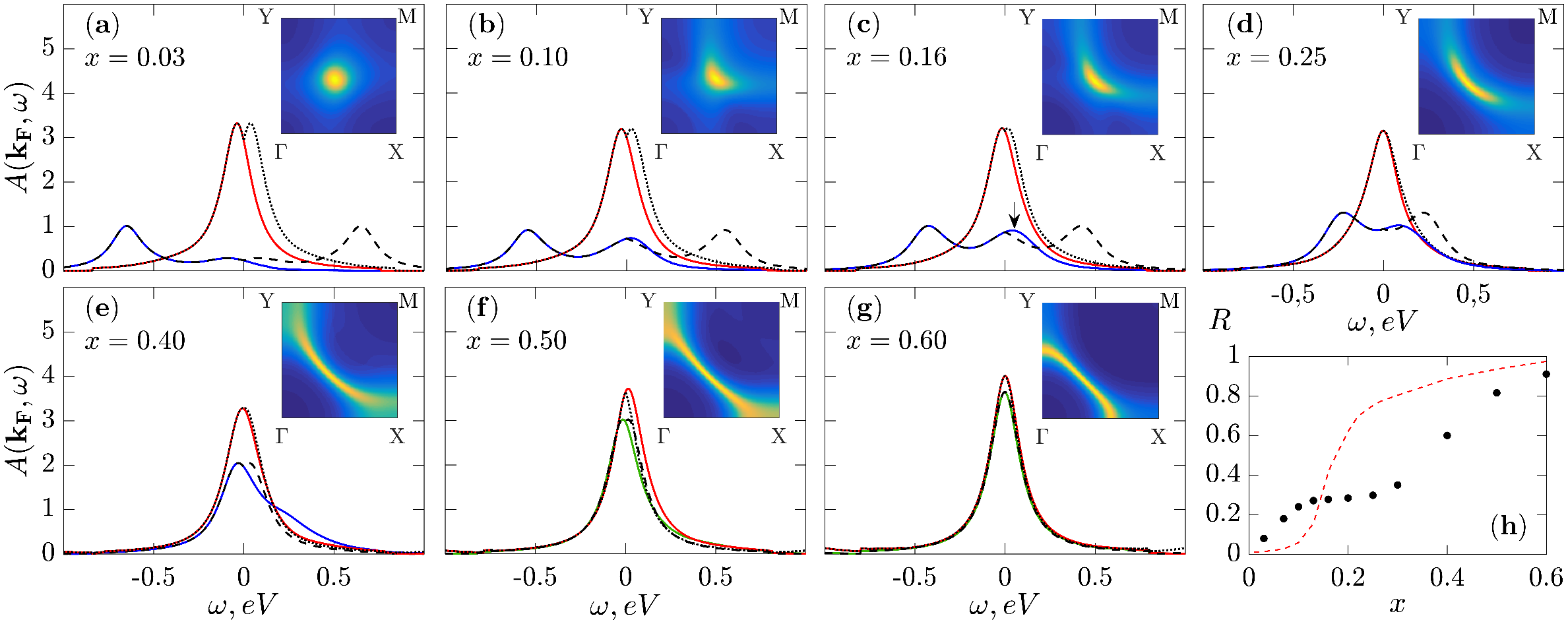}
\caption{\label{fig:f8}
$\lambda_H=1.4$, $\lambda_P=0.02$.}
\end{figure*}
\begin{figure*}
\center
\includegraphics[width=0.95\linewidth]{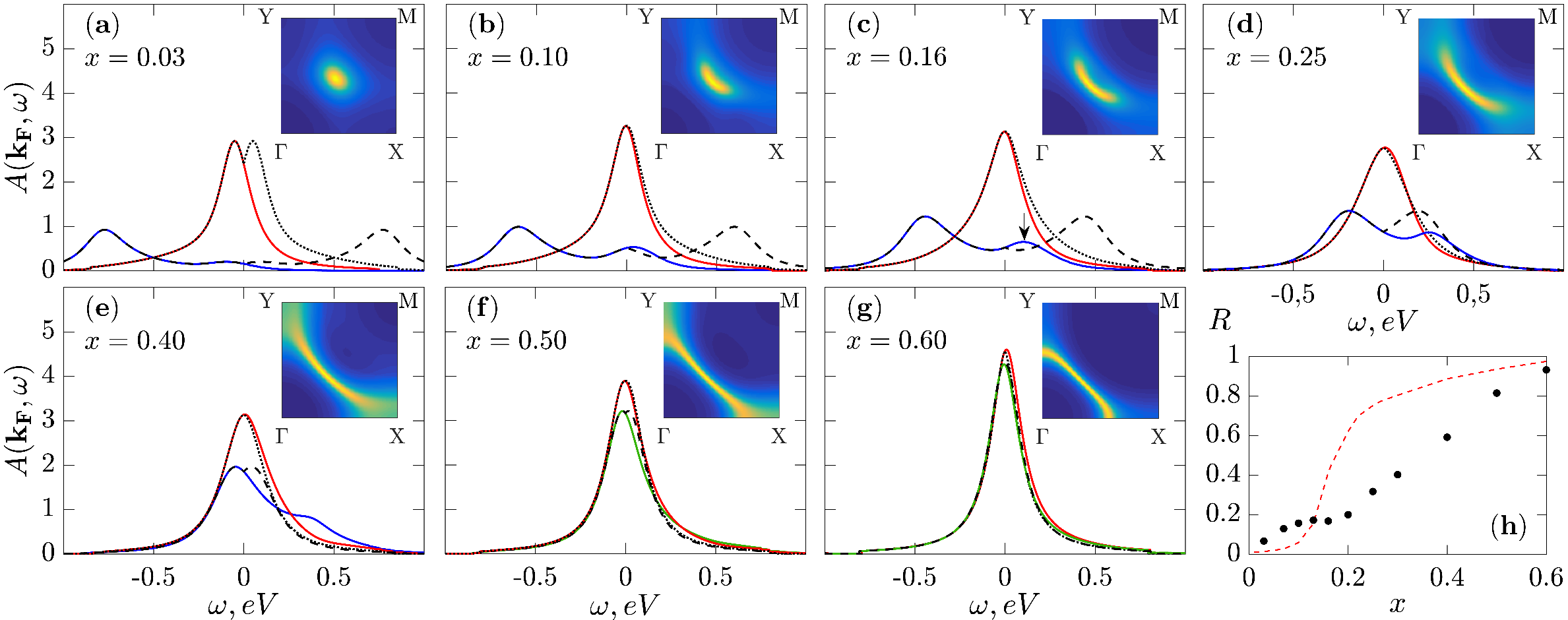}
\caption{\label{fig:f9}
$\lambda_H=1.10$, $\lambda_P=0.03$.}
\end{figure*}
\begin{figure*}
\center
\includegraphics[width=0.95\linewidth]{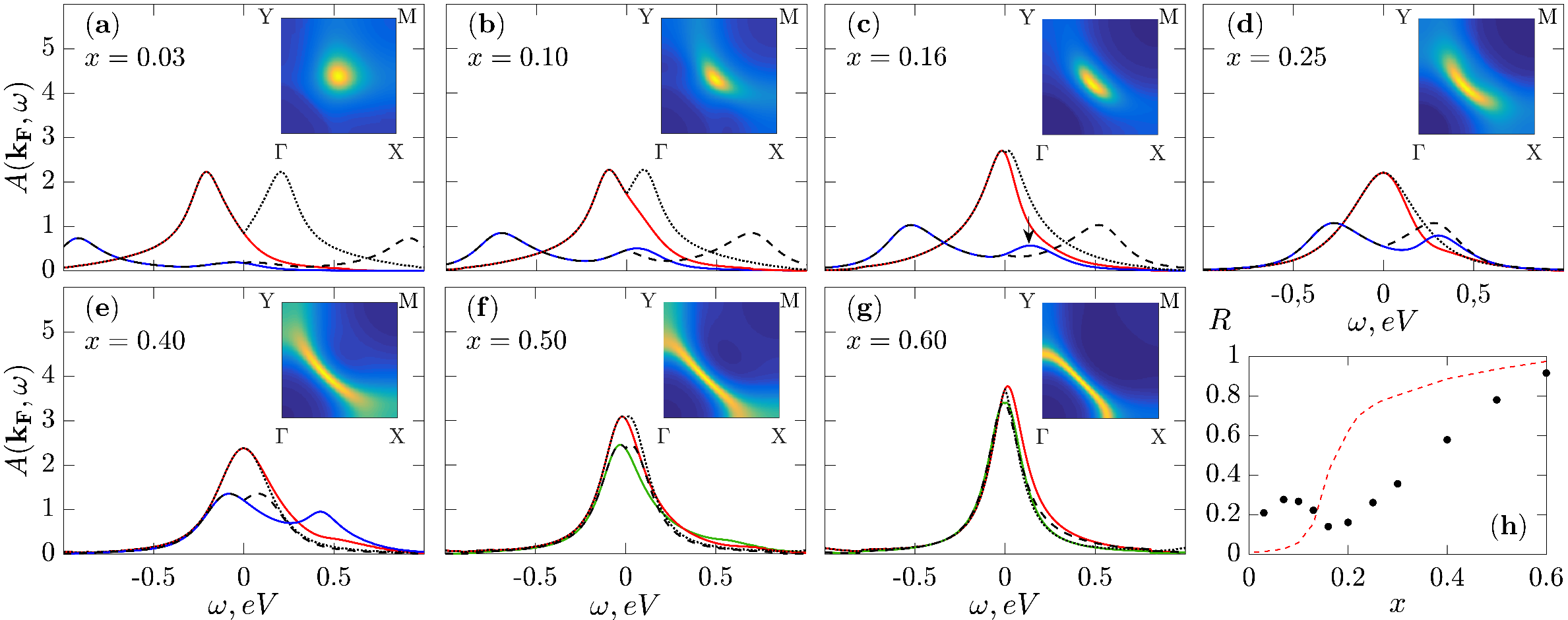}
\caption{\label{fig:f10}
$\lambda_H=1.25$, $\lambda_P=0.03$.}
\end{figure*}
%

%===========================================================================
\section{Special points and regions of the phase diagram \label{sec_PhaseDiagram}}
%===========================================================================
Here, we analyze the conditions that, as a result of the competition between Coulomb and electron-lattice interactions, lead to the formation of specific points and regions on the phase diagram of the system.

%===========================================================================
\subsection{The competition of Coulomb and Holstein interactions \label{subsec_Points}}
%===========================================================================
Let us assume that the main contribution of the EPI in equation~(\ref{H_tot}) arises from the Holstein mechanism, i.e. $M_{pd} =0$ and, consequently,  $\lambda _{P} = 0$. By employing the canonical transformation of Lang and Firsov~\cite{LangFirsov,DudarevSFU}, which allows for the summation of infinite series for the transformed operators, we remove the explicit form of the electron-phonon interaction from the resulting Hamiltonian. The Hamiltonian obtained in this way demonstrates polaron effects in the strong coupling limit. Firstly, this is a change in the local ground state energy ${\tilde{\varepsilon _d}} = \left(\varepsilon_d - \Delta { \langle {n_{{\bf{g},\sigma}}^d} \rangle } \right)$, determined by the polaron shift $\Delta = {\lambda_{H}}W$ and the concentration of charge carriers on the d-orbital $ \langle {n_{{\bf{g},\sigma}}^d} \rangle $. Secondly, the effective decrease of the Coulomb potential by an amount of $ 2 \Delta$. Thirdly, the residual polaron-lattice interaction, which is defined by the renormalization of the creation and annihilation operators of charges on the d-orbitals $\tilde{d}^\dagger_{\mathbf{g},\sigma}=d^\dagger_{\mathbf{g},\sigma} \exp\left[\frac{M^d} {\hbar \omega_0}\left(f^\dagger_{\mathbf{g}}-f_{\mathbf{g}}\right)\right]$. In single-band models, such a contribution leads to polaronic band narrowing due to the reduction of the overlap integral of the orbitals, while in multi-band models with hybridization effects, it also plays a role in the redistribution of electron density among orbitals of different types.

It is well known that in the Holstein model, the transition from a large polaron to a small one, accompanied by a change from the usual to the hopping mobility mechanism, is expected to occur when $\Delta \gg W$. However, in a correlated system characterized by an anisotropic electron band structure and a momentum-dependent bandwidth, quasi-particle excitations lead~\cite{PhysRevB.101.235114,PhysRevB.104.155153} to the formation of a narrow band near the Fermi level, even under the condition $\Delta \approx W$. The critical strength of the electron-phonon coupling that initiates the transformation of the electronic structure can be identified by comparing the polaron shift with the bandwidth, resulting in $\lambda^{\left( { 1 } \right)}_{H} =  {\tilde{W}} / {W \langle {n_{{\bf{g},\sigma}}^d} \rangle} $. Here, we have accounted for the changes in the bandwidth $ {\tilde{W}} $ resulting from the electron-phonon interaction and have explicitly retained the average value of the particle number operator $ \langle {n_{{\bf{g},\sigma}}^d} \rangle $. Due to the redistribution of charge carriers between the $p$ and $d$ orbitals, the value of $\langle {n_{{\bf{g},\sigma}}^d} \rangle$ is not necessarily close to one, even at half-filling.

To determine the second critical point along the $\lambda_{H}$ axis, we compare the on-site energies of the polaron $-\Delta {\langle {n_{{\bf{g},\sigma}}^d} \rangle} $ and the bipolaron $ U_d - 2 \Delta - 2\Delta {\langle {n_{{\bf{g},\sigma}}^d} \rangle} $, shifting the energy scale for convenience so that the condition $\varepsilon _d = 0$ is satisfied. It is evident that in the limit of strong electron correlations, Coulomb repulsion prevents the formation of bipolarons, thereby preserving the dominance of polaron formations. This situation changes with an increase in the strength of electron-phonon coupling. The formation of a local bipolaronic state becomes more favorable since only $-\Delta {\langle {n_{{\bf{g},\sigma}}^d} \rangle} \geq  U_d - 2 \Delta - 2\Delta {\langle {n_{{\bf{g},\sigma}}^d} \rangle} $, from which we immediately obtain that $\lambda^{\left( {2} \right)}_{H} = U_d / W \left( 2+ \langle {n_{{\bf{g},\sigma}}^d} \rangle \right)$.

There is a third critical point on the $\lambda_{H} $-axis, and it corresponds to the condition for the formation of an effective local attractive potential $U_d - 2 \Delta \geq 0$. In previous studies of a single-band model of correlated electrons with strong electron-phonon interactions, it was established that within the range of values $\lambda_{H} \geq \lambda^{\left( {3} \right)}_{H} $, where $\lambda^{\left( {3} \right)}_{H} = U_d / 2W$, the system becomes unstable with respect to the localization of small bipolarons. Note that $\lambda^{\left( {3} \right)}_{H}$ is always greater than $\lambda^{\left( {2} \right)}_{H}$, and in the parameter region $ \lambda_{H} \geq \lambda^{\left( {3} \right)}_{H} $, the electronic structure of the Hamiltonian~(\ref{H_tot}) exhibits~\cite{PhysRevB.104.155153} an insulator state.

The obtained expressions define the sequence of the critical polaron and bipolaron points along the $\lambda_{H}$ axis, governed by both the strength of the electron-phonon coupling and the Coulomb potential. A comparison of the value of $\lambda^{\left( {1} \right)}_{H}$, which determines the onset of polaron effects, with the values of $\lambda^{\left( {2} \right)}_{H}$ and $\lambda^{\left( {3} \right)}_{H}$  demonstrates that the polaron region collapses at $U^{\left( {c} \right)}_d \approx 2W$. For simplicity, here and below, we admit that $\langle {n_{{\bf{g},\sigma}}^d} \rangle \approx 1$. It can be expected that bipolaron formation effects precede polaron ones at lower values of Coulomb repulsion, and vice versa, in the limit of strong electron correlations $U_d > 2W$ and within the regime of intermediate EPI strength from $\lambda^{\left( { 1 } \right)}_{H}$ to $\lambda^{\left( { 2 } \right)}_{H}$, a scenario of electron correlated lattice polarons is realized, that is, effects of polaron transformation of the electronic structure precede bipolaron ones.

%===========================================================================
\subsection{The essential role of Peierls' contribution \label{subsec_PolaronRegion}}
%===========================================================================
Using the above model parameters, we calculated the values of the critical quantities $ U^{\left( {c} \right)}_d $, $\lambda^{\left( { 1 } \right)}_{H} $, and $ \lambda^{\left( { 2 } \right)}_{H}$, with $ \lambda^{\left( { 2 } \right)}_{H}$ also estimated as a function of the Coulomb repulsion parameter $U_d$. The results are shown as colored points along the $\lambda_{H}$ axis of Fig.~\ref{fig:f1}. The same phase diagram demonstrates the lines of polaron (black curve) and bipolaron (colored curves) transformations, obtained through the diagonalization of the cell cluster of the original model, which accounts for both mechanisms of electron-lattice coupling. It is necessary to clarify what we refer to as the lines of polaron and bipolaron transformations. The points on the phase diagram located on the lines of polaron and bipolaron transformations represent special points of functions that characterize the corresponding ground single-particle or two-particle states of the cluster as the strength $ \lambda_{H}$ of the electron-phonon coupling increases. For instance, the critical point $\lambda^{\left( { 1 } \right)}_{H}$, as well as the values of the parameters $\lambda_{H} $ and $\lambda_{P}$ located in the thinner part of the black curve in Fig.~\ref{fig:f1}, specify the inflection points for such functions of $\lambda_{H}$ as the polaron shift, the  mean square displacement, the electron density in the orbitals, and the number of phonons in the ground polaronic state.

It is clear that the analytical estimates based on Lang-Firsov type renormalizations are in good agreement with the numerical calculations. Firstly, the region of the electron correlated lattice polaron on the $\lambda_{H}$ axis, located between points $\lambda^{\left( { 1 } \right)}_{H} $ and $ \lambda^{\left( { 2 } \right)}_{H}$, diminishes with the reduction of the Coulomb repulsion parameter, and its complete collapse occurs precisely at $U_d \approx U^{\left( {c} \right)}_d $. Secondly, the values of the critical quantities $\lambda^{\left( { 1 } \right)}_{H} $ and $ \lambda^{\left( { 2 } \right)}_{H}$ correspond to the onset of the polaron and bipolaron transformation regimes in the parameter space of $\lambda_{H} $ and $\lambda_{P}$ at $\lambda_{P} = 0$.

The electron-phonon coupling of the Peierls type significantly affects the phase diagram. Firstly, starting from a certain value of $\lambda_{P}$, the smooth evolution of the properties of charge carriers at the lines of polaron and bipolaron transformations is replaced by a sharp transitions. In this case, the aforementioned characteristics of the bound states undergo a sudden increase~\cite{PhysRevB.101.235114,PhysRevB.104.155153}. In Fig.~\ref{fig:f1}, the smooth and sharp types of transformations are indicated by thin lines and bold lines, respectively. The non-analytical behavior of polaronic and bipolaronic properties is inherent to models that take into account the transitive mechanism of electron-phonon coupling~\cite{PhysRevB.78.214301,PhysRevLett.105.266605,Sboychakov2010}, while Holstein-type models provide only smooth changes in the ground state~\cite{RevModPhys.63.63}. Quasiparticle excitations that arise from non-perturbative changes in the properties of polarons and bipolarons can significantly influence the characteristics of the system only when they possess sufficient spectral weight. However, the lines of transformations on the phase diagram indicate a shift in the regimes, within which qualitative changes in the properties develop.

The second significant change in the phase diagram in the $\lambda_{H}$ and $\lambda_{P}$ variables is the rapid narrowing of the correlated polaron region with an increase in the strength of the Peierls-type electron-phonon coupling. The intersection of the lines of polaron and bipolaron transformations closes this part of the phase diagram and signals a transition to the bipolaron regime. Thus, the condition for the realization of the correlated polaron regime is not only the relationship $U_d \geq 2W$ but also a sufficiently small Peierls-type electron-phonon contribution, as can be seen from Fig~\ref{fig:f1}. At all points in the phase diagram above the line of bipolaron transformation, we observe ~\cite{PhysRevB.104.155153} an almost complete suppression of the density of states at the Fermi level, which indicates the localization of bound electron-lattice states.

Below, the role of lattice polarons in the formation of a pseudogap will be examined in the range of electron-phonon interaction parameters that describe the system prior to bipolaron transformations. For model~(\ref{H_tot}), this is the region of the phase diagram in Fig.~\ref{fig:f1}, located below the red curve. Within the pGTB approach, we will demonstrate how qualitative changes in the electronic structure evolve at the Fermi level with the varying strength of the electron-lattice interaction and consider their evolution with doping.

%===========================================================================
\section{Analysis of electronic spectra \label{sec_SpectrFunc}}
%===========================================================================
To analyze the pseudogap anomalies in the spectrum, it is convenient to use, along with the usual spectral function, the symmetrized spectral function~\cite{Norman1998}, which is constructed as follows:
\begin{eqnarray}
\label{SymSpectr}
A_{sym} \left( {\bf{k}}_F,\omega \right) & = & A \left( {\bf{k}}_F,\omega \right) f \left( \omega,T \right) + \nonumber \\
&+& A \left( {\bf{k}}_F, -\omega \right) f \left( -\omega,T \right). 
\end{eqnarray}
Here, $ A \left( k_F,\omega \right) $~-- is the spectral function of electrons for the quasi-momentum ${\bf{k}}_F$ on the Fermi surface and the energy $ \omega $, measured from the chemical potential level, while $ f \left( \omega,T \right) $~-- is the Fermi-Dirac distribution function. In the pseudogap state, the symmetrized spectral function $A_{sym} \left( {\bf{k}}_F,\omega \right)$ exhibits a depletion of spectral weight at the Fermi level, which evolves into a quasiparticle peak as the system transitions to Fermi liquid behavior with changes in temperature or doping.

\begin{figure}
\center
\includegraphics[width=\linewidth]{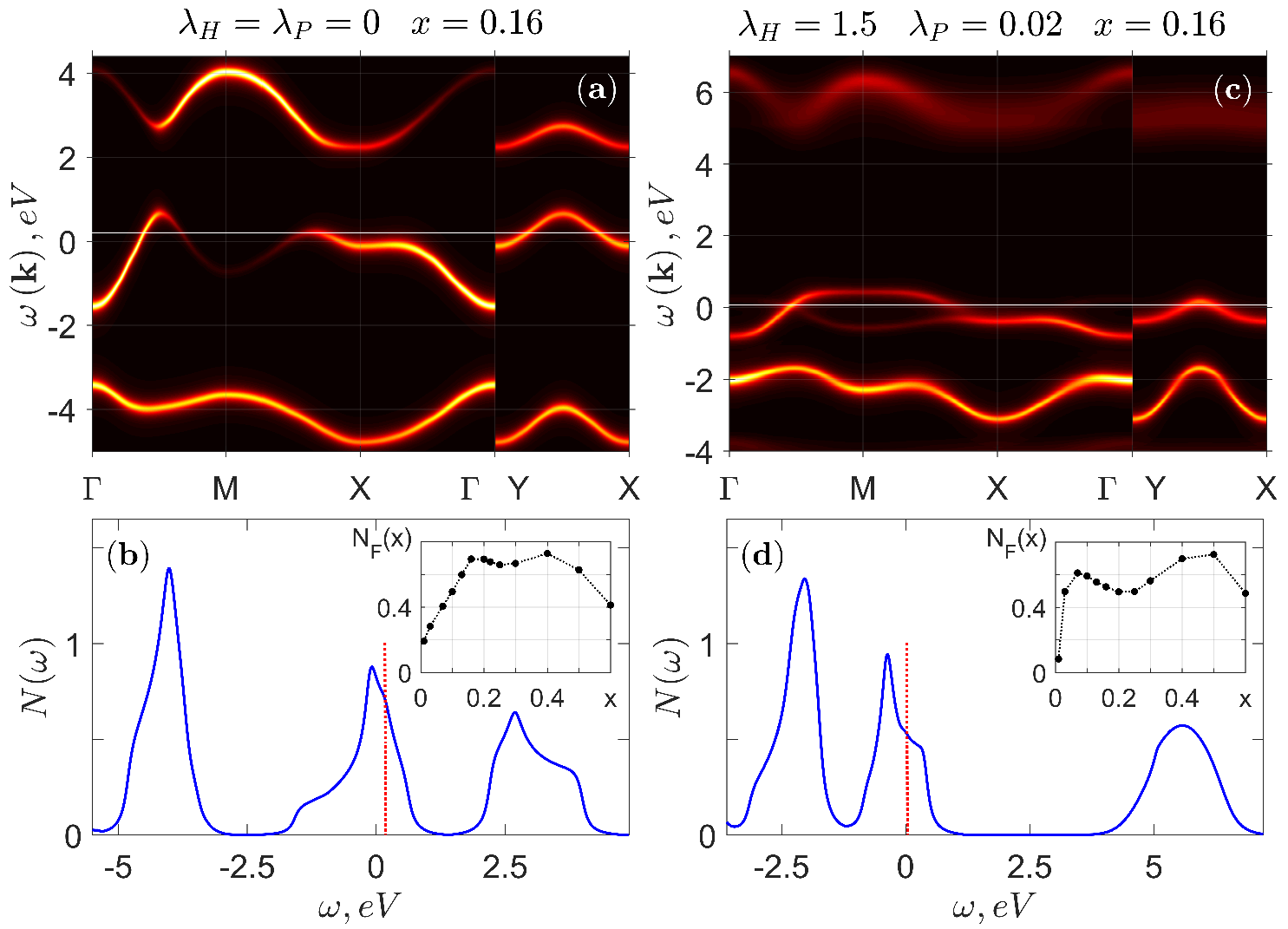}
\caption{\label{fig:f11}
The electron band structure transformation of the system with strong electron correlations as a result of electron-phonon contribution. The dispersion along the main directions of the Brillouin zone $ \omega \left( \bf{k} \right) $ for the parameters as indicated above the figure and the corresponding density of states $N \left( \omega \right)$: (a, b) without considering electron-phonon coupling; (c, d) in the case of strong polaronic effects. The Fermi level is indicated by a thin horizontal line in the upper figures and a vertical red line in the lower ones. The insets demonstrate the density of states at the Fermi level as a function of doping, $N_F \left( x \right)$. At low or optimal concentrations of doped charge carriers $x$, the maximum of this function corresponds to (b) the quantum phase transition that takes place in the doped system with short-range antiferromagnetic order without the involvement of lattice vibrations (the Van Hove singularity), or (d) the change in the topology of the Fermi surface due to the formation of a partially flat band of strongly bound lattice polarons.}
\end{figure}

The spectral functions presented in the series of figures~\ref{fig:f2}~-\ref{fig:f11} are defined for points with maximum spectral weight on the Fermi surface in the nodal $ \Gamma \left( {0,0} \right) \rightarrow M \left({\pi},{\pi }\right)$ (red curve) and antinodal $ X \left( {\pi,0} \right) \rightarrow M \left({\pi},{\pi }\right)$ (blue curve) directions. The antinodal direction $ \Gamma \left( {0,0} \right) \rightarrow X \left({\pi},{0}\right)$ (green curve) is depicted if the spectral function exhibits here greater intensity than in the $\left( X,M \right)$ direction. This occurs when the orientation of pockets or arcs centered on the Fermi surface around the point $\left({\pi},{\pi }\right)$ changes towards the point $\left( 0,0 \right)$ as the concentration of doped carriers increases. The symmetrized spectral functions are shown by corresponding dashed curves. The insets demonstrate the Fermi surface for a given concentration of doped charge carriers $x$. The ratio of the intensities of the spectral functions $R$ characterizes the restoration of the spectral weight in the antinodal direction of the Fermi surface in comparison with the nodal one, $R \left( x \right) =A_{an} \left( {\bf{k}}_F, \omega_F,x \right) / A_{nod} \left( {\bf{k}}_F, \omega_F,x \right)$.

In the limit of strong and intermediate electron correlations, the pseudogap state of a system without electron-phonon interactions emerges against the background of fluctuations in short-range antiferromagnetic order. This leads to an anisotropic rearrangement of the electron band strusture, which evolves with doping from a non-Fermi liquid in lightly doped systems to a typical Fermi liquid in overdoped systems~\cite{Sadovskii_2001,Barabanov2001,Kuchinskii2005,CondMatt.23.045701,PhysRevB.101.115141}, Fig.~\ref{fig:f2}~(a-g). The transition between different regimes occurs through two quantum critical points~\cite{CondMatt.23.045701}, caused by changes in the Fermi surface topology at doping level $x_{c1} \approx 0.18$ and $x_{c2} \approx 0.27$. The first critical point corresponds to the merging of small hole pockets formed near the points $ \left( \pm {\pi }/{2}; \pm {\pi }/{2} \right) $ and is accompanied by a logarithmic Van Hove singularity in the density of states. The second critical point corresponds to the collapse of the inner contour of the Fermi surface around the point $\left({\pi},{\pi }\right)$ and is accompanied by a Heaviside step-type singularity in the density of states.

At low or optimal doping and for $\lambda_P=0$, $\lambda_H=0 $, the symmetrized spectral function demonstrates at the Fermi level $\omega = 0$ a quasiparticle peak in the nodal direction and a dip in the antinodal one, which is characteristic of the pseudogap state, Fig.~\ref{fig:f2}~(a-c). The maximum intensity of the antinodal quasiparticle peak is located below the Fermi surface and is significantly less than the intensity of the nodal peak. With increasing doping, the antinodal quasiparticle peak moves closer to the Fermi surface, which is illustrated by the gradual merging of the peaks of the corresponding symmetrized antinodal spectral function (dashed curve). The complete merging of the peaks occurs at the point of the first quantum phase transition, $x \approx 0.18$. At the next quantum critical point $x \approx 0.27$, the intensity of the antinodal peak begins to rapidly recover, and the system switches into a Fermi liquid regime, Fig.~\ref{fig:f2}~(d-g). The function $R$, which describes the ratio of the intensities of the antinodal and nodal spectral peaks, also exhibits features at the points of quantum phase transitions $x_{c1}$ and  $x_{c2}$. The corresponding changes, specifically the inflection and bending points, are marked in red in Fig.~\ref{fig:f2}~(h).

The evolution of polaron effects with a change in the strength of the electron-lattice coupling and doping is illustrated in Figures~\ref{fig:f3}-\ref{fig:f11}. Let us outline the visual consequences of the interaction between charge carriers and lattice vibrations in the examined region of the phase diagram: (i) a decrease in the spectral intensity of quasiparticle peaks, noticeable even with a small contribution of the electron-phonon interaction (Fig.~\ref{fig:f2} compared to, for example, Figs.~\ref{fig:f3} and~\ref{fig:f4}); (ii) the emergence of a narrow polaron band in the high-energy part of the spectrum at intermediate coupling strength (for clarity, the corresponding peak is indicated with a small black arrow in some figures); (iii) pronounced asymmetry of the nodal quasiparticle peak in the limit of strong electron-phonon interaction (Fig.~\ref{fig:f2} versus Figs.~\ref{fig:f4} and~\ref{fig:f6}).

The reduction in spectral intensity exhibits several intriguing patterns in the parameter region preceding the aforementioned asymmetry of the spectral function. Firstly, regardless of the type of electron-phonon coupling mechanisms or the directions of the spectrum considered, a stronger suppression of quasiparticle peaks is observed in the non-Fermi liquid regime, meaning before the restoration of the large Fermi surface, which is characteristic of free charge carriers. Secondly, when considering only the Holstein contribution, the decrease in intensity is predominant in the antinodal direction. The described modulation of the spectral intensity enhances the blurring of the shadow part of the antiferromagnetic pockets that are closer to the point $\left({\pi},{\pi }\right)$. However, qualitative changes in the electronic structure do not occur (Fig.~\ref{fig:f3}, ~\ref{fig:f5} versus Fig.~\ref{fig:f2}) until the top of the valence band of electrons (or the conduction band of holes) located near the Fermi level is transformed by the coherent polaron excitations.

The formation of a narrow, partially flat polaron band begins around the point $\left({\pi},{\pi }\right)$  and is accompanied by the emergence of an additional peak with lower intensity in the antinodal direction of the spectral function. In order to understand how the parameters of electron-lattice coupling and doping control the effects of the narrow polaron band, we performed a detailed study of the phase diagram of the system (see Appendix~\ref{sec:a1}).

It was found that (i) the narrow polaron band aligns with the Fermi level in the region of the EPI parameters describing the correlated polaron, or approaching this region  (that is, near and above the black line corresponding to the polaron, but below the red line indicating the bipolaron transformations, to the left of the point of their intersection in Fig.~\ref{fig:f1}); (ii) this occurs in the non-Fermi-liquid regime at concentrations of doped holes that are less than or approximately equal to the first critical value $x_{c1}$.  

The examples of the evolution of the spectral function for the corresponding set of parameters $\lambda_H$ and $\lambda_P$ are shown in figures~\ref{fig:f7}-\ref{fig:f10}. It is evident that the emergence of the narrow and partially flat polaron band at the Fermi level opens up the small antiferromagnetic pocket, transforming it into an arc (Fig.~\ref{fig:f5}a-e versus Figs.~\ref{fig:f7}b-c,~\ref{fig:f8}a-d,~\ref{fig:f9}a-d and~\ref{fig:f10}a-d). We emphasize that the reduction in the intensity of the shadow part of the hole pockets near the $\left({\pi},{\pi }\right)$ point, as observed in Figures~\ref{fig:f5}a-e, is associated with the damping of spin correlation functions as doping increases. While Figures~\ref{fig:f7}-\ref{fig:f10} demonstrate the change in the topology of the Fermi surface due to the formation of a narrow polaron band near the Fermi level. With an increasing concentration of doped particles, the arcs grow, evolving from small arcs centered around the $\left({\pi},{\pi }\right)$ point to a large Fermi surface centered around the $\left({0},{0}\right)$ point (Figs.~\ref{fig:f8}-\ref{fig:f10}). If the narrow polaron band has a small spectral weight, it quickly shifts away from the Fermi level as doping increases. In this case, one can observe a restoration of the band structure evolution that is characteristic of a system with zero or weak EPI contributions (Figs.~\ref{fig:f7}d,e and Figs.~\ref{fig:f2}d,~\ref{fig:f5}d,e). The doping range in which the narrow polaron band determines the shape and new topology of the Fermi surface is clearly demonstrated by the intensity ratio function $ R \left( x \right) $. Indeed, the introduction of electron-lattice coupling causes the features of this function related to the changes of the Fermi surface at doping levels $x_{c1}$ and $x_{c2}$ to become progressively less pronounced (Fig.~\ref{fig:f2}h versus Figs.~\ref{fig:f5}h and ~\ref{fig:f3}h). At the same time, new features emerge, a local maximum and minimum, resulting from changes in the density of states due to the formation of the narrow polaron band at the Fermi level (Figs.~\ref{fig:f7}h-\ref{fig:f10}h).

Qualitative changes in the Fermi surface resulting from polaron effects are clarified in Figure~\ref{fig:f11}. The insets (a) and (c) present a comparison between the band structure of a non-modulated system and a crucial stage of a strongly modulated one. In the system characterized by strong EPI, the partially flat polaron band, formed around the point $\left({\pi},{\pi }\right)$, overlaps with the original top of the valence band lying below it at the points $\left( {\pm \pi }/{2};{\pm \pi }/{2} \right)$. This overlapping further extends to the original band structure near the $X$ and $Y$. If the Fermi level lies within the energy range from the top of the newly formed hybrid band to the level corresponding to the top of the untransformed band, an arc-like Fermi surface is observed. In the corresponding part of the spectrum, the density of states $N \left( \omega \right)$ exhibits a feature in the form of a shoulder, which leads to the maximum of the function $N_F \left( x \right)$.

%===========================================================================
\section{Conclusion and discussion of results \label{sec_Discussion}}
%===========================================================================
The presented results demonstrate that in a system with strong electron correlations, the interaction of charge carriers with lattice vibrations can control both the quantitative modulation of the spectral characteristics of the pseudogap state and the emergence of a new pseudogap mechanism. The transformation of the electronic structure, driven by the development of a partially flat polaron band near the Fermi level, is accompanied by a suppression of the intensity of quasiparticle excitations that undergo scattering due to the spin fluctuations of the short-range antiferromagnetic order, as well as the formation of a top of the valence band of electrons around the $\left( {\pi},{\pi} \right)$ point. The revealed features may indicate a suppression of short-range magnetic order due to excitations of strongly bound electron-lattice quasiparticles.

In general, the processes of virtual quasiparticle transitions responsible for the effective exchange interaction can occur in two ways~\cite{Kugel1980}. If the lattice relaxes during the intermediate state, a significant reduction in the effective exchange integrals is observed, explained by the exponential suppression of the hopping integrals. In the opposite limit, the virtual transition occurs according to the Franck-Condon principle for a frozen lattice, and the exchange integral is not renormalized. The above studies show that in a system with electron correlations and an anisotropic electron band structure, strong electron-phonon interactions can lead to competition between lattice relaxation processes which are “fast” and “slow” in comparison to the exchange interaction.

The destruction of the short-range antiferromagnetic order with doping, as previously described for correlated systems~\cite{Sadovskii_2001,Barabanov2001,Kuchinskii2005,CondMatt.23.045701,PhysRevB.101.115141}, differs from the case presented here not only in its nature. Firstly, the formation of the flat band section near the Fermi level and the change in the topology of the Fermi surface occur only within a specific region of the phase diagram and can be observed, including as a feature in the density of states, even at low levels of doping. Secondly, the processes of transformation of the electron band structure induced by excitations of strongly bound lattice polarons are accompanied by renormalizations of the phonon spectrum. The corresponding dispersion and shift of the bare phonon line away from its resonance value were demonstrated earlier in paper~\cite{PhysRevB.101.235114}.

The findings reported here contribute to a deeper understanding of the intriguing phase diagrams of systems characterized by complex interplay between electron-electron and electron-phonon interactions, including high-temperature superconductors.

\begin{acknowledgments}
The research presented in sections \ref{intro}-\ref{sec_PhaseDiagram} was carried out within the state assignment of the Kirensky Institute of Physics. The study reported in section \ref{sec_SpectrFunc}-\ref{sec_Discussion}  has been supported by Russian Science Foundation (RSF project No. 24-12-00044).
\end{acknowledgments}

%%%%%%%%%%%%%%%%%%%%%%%%%%%%%%%%%%%%%%%%%%%%%%%%%
\appendix
\section{\label{sec:a1} Principles and Practical Implementation of the pGTB Method}
The polaronic version of the generalized tight-binding method (p-GTB), as well as the original GTB, belongs to the cluster perturbation theory framework. The approach begins with the decomposition of the full Hamiltonian~(\ref{H_tot}) into intracluster and intercluster components $H = H_c + H_{cc}$, where the intracluster term $H_c = \sum\nolimits_{\bf{g}} {H_{\bf{g}}}$ is summed over all lattice sites ${\bf{g}}$. The problem is addressed via canonical transformations of the Hamiltonian, with the transformation matrices depending on the unit cell symmetry, the irreducible representations of its point group, and the symmetry properties of the phonon modes~\cite{PhysRevB.53.8751,PhysRevB.53.8774,PhysRevB.59.14697,Gav2000,Shneyder18}. The constructed unit cell cluster captures the realistic structure and chemical bonding environment of the planar tetragonal system, while the incorporated full-breathing mode preserves its point symmetry.

In the new symmetric cell representation, the strong on-site localization of Wannier orbitals induces the renormalization of all parameters of the full Hamiltonian. For example, the parameters $ U_p $ and $ V_{pd} $ of the Coulomb interactions related to ligand holes transform as ${U_p}{\Psi _{gfmn}}$ and $ V_{pd}{\Phi _{gfm}}$, respectively. The intracell coefficients are $ \Psi _{0000} = 0.21$ and $ \Phi _{000} = 0.918$, while the intercell terms are $ \Psi _{0001} = -0.03$, $ \Phi _{001} = 0.13$, and $ \Phi _{002} = -0.02$. Thus, the Coulomb interactions are strongly localized within the unit cell of $H_{\bf{g}}$, allowing the intercell contributions from $U_p$ and $V_{pd}$ to be neglected. Generally, this approach introduces a methodological constraint: the retained intercluster electron-phonon contributions must dominate over the discarded intercluster Coulomb interaction terms.

The renormalized electron-phonon coupling parameters $M_{{\bf{gg}}'}^d = {M_d}{\mu _{{\bf{gg}}'}}$ and $M_{{\bf{g}\bf{g}'\bf{h}}}^{bd} =  - 2{M_{pd}}{\mu _{{\bf{gg}}'}}{\mu _{{\bf{hg}}'}}$ also exhibit rapid distance decay, as quantified by $\mu _{00} = 0.96$, $\mu _{01} = -0.14$, and $\mu _{11} = -0.02$. Despite their short-range nature, these intercluster contributions remain essential in certain physical contexts, for example, in determining the dispersion of phonon excitations~\cite{PhysRevB.101.235114}. Furthermore, these terms restore the full Hamiltonian's translational symmetry when modified by a vibrational phase factor accounting for inter-site phase differences. To date, the simultaneous accurate treatment of both long-range and short-range correlations in systems with strong interactions remains an unsolved problem. We assume that in the strong-coupling regime, the key features of small-polaron electronic structure and the density of states are predominantly determined by the competition between localized electron-electron and electron-phonon interactions. Consequently, the aforementioned intercluster terms, including phase correlations, have been neglected in our analysis.

The complete expression for the Hamiltonian~(\ref{H_tot}) in its intermediate representation, $H_c + H_{cc}$, is given in Ref.~\cite{PhysRevB.101.235114}. The next step involves the exact diagonalization of the effective Hamiltonian for the unit cell cluster described by the Hamiltonian $H_{\bf{g}}$. The procedure is performed for each system configuration characterized by $n_h$ holes per unit cell and followed by a controlled truncation of high-energy states with minimal spectral weight. Our truncation of the phonon Hilbert space ensures that the ground state and first excited states energies converge with less than $1\% $ error for all cases (0, 1, and 2 holes per site) while preserving all essential spectral features.

Using the complete set of multielectron-multiphonon eigenstates  $ {\left| q \right\rangle } $ obtained through exact diagonalization of $H_{\mathbf{g}}$, we express the full Hamiltonian in terms of Hubbard X-operators. Indeed, any operator $A_{\bf{g}}$ can be presented as a linear combination:
\begin{eqnarray}
 \label{O_inX}
A_{\bf{g}} = \sum\limits_{q,q'} {\left\langle q \right|A_{\bf{g}}\left| {q'} \right\rangle {X_{\bf{g}}^{q,q'}}}  = \sum\limits_{Q} {{\gamma}_{A,Q} {X_{\bf{g}}^{Q}}},
\end{eqnarray}
where the index $q$ encompasses all relevant quantum numbers and ${\gamma _{A,Q}} = \left\langle {q\left| {{A_{\bf{g}}}} \right|q'} \right\rangle $ is the matrix element for the transition from the initial state $\left| q' \right\rangle $ to the final state $\left| q \right\rangle $, i.e. for a pair of states $Q = \left( {q,q'} \right)$, under the action of the operator $A$. The Hamiltonian then reduces to the compact form:
\begin{align}
\label{X_Hc}
&H_c = \sum\limits_{{\bf{g}},q} {{E_{q}}{X_{\bf{g}}^{q,q}}}, \\
&{H_{cc}} = \sum\limits_{ {{\bf{g}}\ne{\bf{g}}'\ne{\bf{g''}}} } {\sum\limits_{Q,Q'} {\left( {T_{{\bf{gg'}}}^{Q'Q}{\delta _{{\bf{g''g'}}}}\overset{\dagger }{\mathop{X_{\mathbf{g}}^{Q'}}}X_{{\bf{g'}}}^{Q} + } \right.} }  \nonumber \\
\label{X_Hcc}
& + \sum\limits_{Q''} {\left( {M_{{\bf{gg'}}}^{Q''Q'}{\delta _{{\bf{g''g'}}}}{\delta _{Q'Q}} + M_{{\bf{gg'}}{\bf{g''}}}^{Q''Q'Q}} \right)\left. {X_{\bf{g}}^{Q''}\overset{\dagger }{\mathop{X_{\mathbf{g'}}^{Q'}}}X_{\bf{g''}}^{Q}} \right)}, 
\end{align}
with coefficients as 
\begin{align*}
&T_{{\bf{gg'}}}^{Q'Q} = {\tilde{t}}_{{\bf{gg'}}}^{pp}\gamma _{{\tilde{p}}_\sigma ,Q'}^\dag {\gamma _{{{\tilde{p}}_\sigma },Q}} + {\tilde{t}}_{{\bf{gg'}}}^{pd}\Gamma _{Q'Q}^{pd}, \\
&M_{{\bf{gg'}}}^{Q''Q'} = {\tilde{M}}_{{\bf{gg'}}}^d\Gamma _{Q''}^{\tilde{f}}\gamma _{{d_\sigma },Q''}^\dag {\gamma _{{d_\sigma },Q'}},\\  
&M_{{\bf{gg'}}{\bf{g''}}}^{Q''Q'Q } = {\tilde{M}}_{{\bf{gg''g'}}}^{pd}\Gamma _{Q''}^{\tilde{f}}\Gamma _{Q'Q}^{pd}, \\
&\Gamma _{Q'Q}^{pd} = \gamma _{d_\sigma ,Q'}^\dag {\gamma _{{{\tilde{p}}_\sigma },Q}} + \gamma _{{\tilde{p}}_\sigma ,Q'}^\dag{\gamma _{{d_\sigma },Q}}, \\
&\Gamma _{Q''}^{\tilde{f}} = \gamma _{{\tilde{f}},Q''}^\dag  + {\gamma _{\tilde{f},Q''}}.
\end{align*}
Here, ${E_{q}}$ denotes the eigenenergy of state $\left| q \right\rangle $. All operators and parameters resulting from canonical transformations of the Hamiltonian~(\ref{H_tot}) are marked with a tilde. The renormalized parameters take the following form: ${\tilde{t}}_{{\bf{gg'}}}^{pd} =  - 2{t_{pd}}{\mu _{{\bf{gg'}}}}$, ${\tilde{t}}_{{\bf{gg'}}}^{pp} =  - 2{t_{pp}}{\nu _{{\bf{gg'}}}}$, ${\tilde{M}}_{{\bf{gg}}'}^d = {M_d}{\mu _{{\bf{gg'}}}}$, and ${\tilde{M}}_{{\bf{g}\bf{g'}\bf{g''}}}^{pd} =  - 2{M_{pd}}{\mu _{\bf{gg'}}}{\mu _{{\bf{g''g'}}}}$. The coefficients ${\mu _{{\bf{gg'}}}}$ and ${\nu _{{\bf{gg'}}}}$ are Fourier transforms of ${\mu _{\bf{k}}} = \sqrt {s_{x,{\bf{k}}}^2 + s_{y,{\bf{k}}}^2} $ and  ${\nu _{\bf{k}}} = {{4s_{x,{\bf{k}}}^2s_{y,{\bf{k}}}^2} \mathord{\left/ {\vphantom {{4s_{x,{\bf{k}}}^2s_{y,{\bf{k}}}^2} {\mu _{\bf{k}}^2}}} \right. \kern-\nulldelimiterspace} {\mu _{\bf{k}}^2}}$, respectively, with ${s_{x\left( y \right),{\bf{k}}}} = \sin \left( {{{{k_{x\left( y \right)}}{a_{x\left( y \right)}}} \mathord{\left/ {\vphantom {{{k_{x\left( y \right)}}{a_{x\left( y \right)}}} 2}} \right. \kern-\nulldelimiterspace} 2}} \right)$ and ${a_{x\left( y \right)}}$ representing the lattice parameter.

It should be emphasized that the employed mathematical formalism accurately captures the essential physics of band structure formation in correlated systems, particularly the spectral weight redistribution among many-body quasiparticle excitations arising from the local competition between strong electron-electron and electron-lattice interactions. The intensity of these quasiparticle excitations is governed by the overlap of the matrix elements of the participating states and their occupation numbers. Consequently, the redistribution of spectral weights between quasiparticle bands depends on doping, temperature, and the relative strengths of electron-phonon coupling and Coulomb interactions.

Within the Hubbard operator approach, the retarded two-time single-particle Green's function expands into a series of quasiparticle propagators
\begin{equation}
\begin{aligned}
\label{ElGrX}
 & G_{\mathbf{gg'},\sigma}^{\alpha \alpha '}\left( t,t' \right) = \\
& \sum\limits_{Q_h,Q_{h'}}{{{\gamma }_{{{a}_{\alpha ,\sigma }},Q_h}}\gamma _{{{a}_{\alpha '\sigma }},Q_{h'}}^{\dagger }
\left\langle {\left\langle {{X_{{\bf{g}},\sigma }^{{Q_h}}\left( t \right)}}
 \mathrel{\left | {\vphantom {{\overset{\dagger } {X}{}_{{\bf{g}},\sigma }^{{Q_h'}}\left( t \right)} {X_{{\bf{g'}},\sigma }^{{Q_{h'}}}\left( {t'} \right)}}}
 \right. \kern-\nulldelimiterspace}
 {\overset{\dagger}{X}{}_{{\bf{g'}},\sigma }^{{Q_{h'}}}\left( {t'} \right)} \right\rangle } \right\rangle },
\end{aligned}
\end{equation}
where $a_{\alpha ,\mathbf{g},\sigma }$ represents the annihilation operator for a hole with spin $\sigma$ at a lattice site $\mathbf{g}$ on the orbital $\alpha$ and the indices $Q_h$ enumerate all allowed transitions between pairs of eigenstates where the hole number decreases by one in the final state.

To determine the quasiparticle band structure of correlated holes coupled to phonons, we employ~\cite{PhysRevB.101.235114} the projection operator method in the equation of motion for the matrix Green's function $\hat{G}_{\mathbf{gg'},\sigma}^{{{Q}_{h}}{{Q}_{h'}}}\left( t,t' \right)$. In the generalized mean-field approximation, which incorporates the interaction of charge carriers with spin fluctuations, the quasiparticle spectrum is described by the matrix ${{{\hat{\omega }}}_{\mathbf{k},\sigma }}^{{{Q}_{h}}{{Q}_{h'}}}$, with elements given by 
\begin{eqnarray}
\label{elspectr}
\omega_{\mathbf{k},\sigma}^{{{Q}_{h}}{{Q}_{h'}}}=F_{H}^{{{Q}_{h}}}F_{H}^{{{Q}_{h'}}}T_{\mathbf{k}}^{{{Q}_{h}}{{Q}_{h'}}}\pm 
\sum\limits_{\mathbf{q}}{{{c}_{\mathbf{q}}}T_{\mathbf{k}-\mathbf{q}}^{{{Q}_{h}}{{Q}_{h'}}}}\\ \nonumber
+F_{H}^{{{Q}_{h}}}{{E}_{{{Q}_{h}}}}{{\delta }_{{{Q}_{h}},{{Q}_{h'}}}}.
\end{eqnarray}
The plus (minus) sign is taken when the transitions $Q_h$ and $Q_{h'}$ belong to the same (different) subspaces of the Hilbert space, ${\delta }_{{{Q}_{h}},{{Q}_{h'}}}$ is the Kronecker delta-symbol, $F_{H}^{{{Q}_{h}}} = \left\langle { {X}_{\mathbf{g}}^{q,q} + {X}_{\mathbf{g}}^{q',q'} } \right\rangle$ for index ${Q}_{h} = \left({q,q'}\right)$, and ${E}_{{Q}_{h}} = \left( {E_{q'} - E_{q}} \right)$. Fourier transform of the static spin correlation function is given by expression ${c_{\bf{q}}} = \sum\limits_{\left( {{\bf{g}} - {\bf{g}}'} \right)} {{c_{{\bf{gg}}'}}\exp \left[ { - {\bf{q}}\left( {{\bf{g}} - {\bf{g}}'} \right)} \right]} $, where ${c_{{\bf{gg}}'}} = 3\left\langle {S_{\bf{g}}^zS_{{\bf{g}}'}^z} \right\rangle$. It characterizes~\cite{Korshunov2007} the doping-dependent evolution of spin-liquid properties of underdoped cuprates. 

The present implementation of the method correctly describes short-range correlations of competing electron-electron and electron-phonon interactions. This enables analysis of small-polaron electronic structure evolution in the normal state. However, to describe phenomena such as superconducting states, large-to-small polaron transitions, or large polaron properties, the inclusion of relevant long-range correlations is required.

To systematically explore the phase diagram of the system, we calculate the electronic band structure using adaptive resolution for electron-phonon couplings ($\lambda_H$, $\lambda_P$) and variable doping intervals $\Delta x$. In parameter-sensitive regions where spectral properties exhibit substantial variations, we employ minimum step sizes of $0.05$ for $\lambda_H$ and $0.005$ for $\lambda_P$, with a doping concentration resolution of $0.01$. Specifically, for the system with $U_d = 9$ (defined in Section~\ref{sec_model}), we conducted a detailed investigation of the parameter space region $0.8 \leq \lambda_H \leq 1.6$ and $0 \leq \lambda_P \leq 0.08$. For each parameter combination, we recompute the complete electronic band structure, including the chemical potential. Our pGTB simulations adopted an  $400\times 400$ k-point mesh for full Brillouin zone sampling. All calculations were performed on high-performance computing clusters using parallelized algorithms. For the parameter range considered in Section~\ref{sec_SpectrFunc}, the average execution time was $15$ minutes per phase diagram point. We note that the computational cost increases substantially with decreasing phonon frequency $\omega_{ph}$ or as the electron-phonon coupling $\lambda_H$ approaches its critical value $\lambda_H^{\left( 3 \right)} $. In the present work, we do not examine the band structure in these parameter regimes. Furthermore, our analysis does not include interactions with Jahn-Teller active vibrational modes that lift the orbital degeneracy of ground electronic states, nor temperature-dependent effects, which may become significant~\cite{Shneyder18,Kugel1980} in polaron systems due to Frank-Condon resonance excitations.
%%%%%%%%%%%%%%%%%%%%%%%%%%%%%%%%%%%%%%%%%%%%%%%%%

\bibliography{mybibfileDA}
\end{document}